\documentclass[12pt,a4paper]{article}

\usepackage{a4}
\usepackage{amsfonts}
\usepackage{amssymb}
\usepackage{epsfig}
\usepackage{psfig}
\usepackage{psfrag}
\usepackage{dsfont}

\newcommand{\beqn}{\begin{eqnarray}}
\newcommand{\eeqn}{\end{eqnarray}}
\newcommand{\be}{\begin{equation}}
\newcommand{\ee}{\end{equation}}
\newcommand{\non}{\nonumber \\}

\def\s1{$s_{\alpha}$}
\def\s2{$s_{\gamma}$}
\def\s3{$s_{\delta}$}
\def\c1{$c_{\alpha}$}
\def\c2{$c_{\gamma}$}
\def\c3{$c_{\delta}$}

\newcommand{\cn}{{\cal N}}

\newcommand{\cv}{{\cal V}}

\newcommand{\co}{{\cal O}}
\newcommand{\ck}{{\cal K}}
\newcommand{\tr}{{\rm tr}}

\begin{document}
\thispagestyle{empty}

\begin{flushright}
\vspace{-3cm}
{\small 
NUB-TH-3252 \\
MIT-CTP-3567 \\[0cm]
hep-th/0411201
}
\end{flushright}
\vspace{1cm}

\begin{center}
{\Large\bf
Hierarchically Split Supersymmetry \\[.15cm]  
with Fayet-Iliopoulos D-terms \\[.35cm]
in String Theory}
\end{center}

\vspace{1.0cm}
\begin{center}

{\bf Boris K\"ors}\footnote{e-mail: kors@lns.mit.edu}$^{,*}$
{\bf and Pran Nath}\footnote{e-mail: nath@neu.edu}$^{,\dag}$
\vspace{.5cm}

\hbox{
\parbox{8cm}{
\begin{center}
{\it
$^*$Center for Theoretical Physics \\ 
Laboratory for Nuclear Science \\ 
and Department of Physics \\ 
Massachusetts Institute of Technology \\ 
Cambridge, Massachusetts 02139, USA \\
}
\end{center}
} 
\hspace{-.5cm}
\parbox{8cm}{\begin{center}
{\it
$^\dag$Department of Physics \\ 
Northeastern University \\
Boston, Massachusetts 02115, \\ USA \\
}
\end{center}
}
}
\vspace{.5cm}
\end{center}

\begin{center}
{\bf Abstract} \\
\end{center}
We show that in string theory or supergravity with supersymmetry 
breaking through combined F-terms and Fayet-Iliopoulos D-terms, the masses for 
charged scalars and fermions can be hierarchically split. The mass scale for 
the gauginos and higgsinos of the MSSM is 
controlled by the gravitino mass $m_{3/2}$, as usual, 
while the scalars get extra contributions from the D-terms of extra abelian $U(1)$ factors, 
which can make them much heavier. The vanishing of the vacuum 
energy requires that their masses lie below $\sqrt{m_{3/2} M_{\rm Pl}}$, 
which for $m_{3/2}=\co({\rm TeV})$ sets a bound of $10^{10-13}\, {\rm GeV}$. 
Thus, scalars with non-vanishing $U(1)$ charges typically become heavy, while 
others remain light, producing a spectrum of scalars
with masses proportional to their charges, and therefore non-universal. 
This is a modification of the split supersymmetry scenario, but with a light
gravitino. We discuss how Fayet-Iliopoulos terms of this size can arise in orientifold 
string compactifications with 
D-branes. Furthermore, within the frame work of D-term inflation, the same vacuum 
energy that generates the heavy scalar masses can be responsible for driving 
cosmological inflation. 
\clearpage
\setcounter{footnote}{0}


\section{Introduction} 

The conventional approach to supersymmetry breaking in models of supergravity (SUGRA) 
is to assume some form of spontaneous breaking in a hidden sector, mediated 
to the visible sector with contains the MSSM via gravitational interactions \cite{Chamseddine:1982jx}. 
The overall mass scale is then set by the gravitino mass $m_{3/2}$, and all other masses 
for squarks, sleptons, gauginos and higgsinos come out roughly proportional to it, 
by demanding the cancellation of the vacuum energy. 
Low energy supersymmetry, as required by a natural explanation of the Higgs potential, 
fixes $m_{3/2}$ to the electro-weak scale. Now, recently it was argued \cite{Arkani-Hamed:2004fb} that 
the fine tuning problem of the Higgs mass may be insignificant compared to the 
fine tuning of the cosmological constant, and that an anthropic selection mechanism 
(see e.g.\ \cite{landscape}) 
may then involve actual fine tuning of MSSM parameters. 
The mass pattern that was proposed under the name of 
split supersymmetry \cite{Giudice:2004tc} has  
all the MSSM fermions at the electro-weak 
scale, whereas all scalars, except for the one fine tuned Higgs doublet, 
get ultra-heavy at a high mass scale. This scenario has attracted some 
attention recently \cite{recentwork}. More concretely, the challenge for model building is to 
keep the gauginos and higgsinos light, while letting the scalars become very heavy. 
The motivation for this originates from supersymmetric grand unification, even 
without low energy supersymmetry in the usual sense, and the model is designed to keep 
the merits of gauge coupling unification as in the MSSM. 
Here, we pursue the perspectives of such patterns in the mass spectrum in the context of 
SUGRA and string theory models, based on the paradigm of spontaneous breaking 
in a hidden sector. 
\\

Given the above mentioned relations that govern gravity mediated supersymmetry breaking, 
it seems very hard to achieve hierarchically split mass scales. If all masses are proportional 
to $m_{3/2}$, there is no room for flexibility. There is however a loophole to the argument, 
which has not so far been explored in the conventional approach in any depth, probably because 
it quickly leads to large masses, which were thought unacceptable. It consists of 
assuming not only auxiliary F-terms but also D-terms to be generated. In global 
supersymmetry, this is well known under the label of supersymmetry breaking mediated 
by an anomalous $U(1)$, where large masses can be 
avoided \cite{Dvali:1996rj,Binetruy:1996uv}, as will be seen later.
The mechanism basically adds a Fayet-Iliopoulos (FI) term for an extra 
anomalous $U(1)$ gauge symmetry, independent of the F-terms which may also 
be present. In local supergravity, the two contribute both to the vacuum energy, and thus get tied 
together at roughly the same scale. Still, the contribution to scalar masses can 
be very different, since the F-terms are mediated via gravity, while the D-terms are 
mediated via gauge interactions, which opens up the possibility to have 
hierarchically split mass scales, splitting scalars charged under the relevant $U(1)$, 
from all other fields, i.e.\ gauginos, higgsinos as well as scalars not charged under 
the relevant $U(1)$'s. 
This is not quite along the lines of high scale supersymmetry breaking, as advocated in 
split supersymmetry, where the gravitino mass itself was assumed at 
the high scale, and the main difficulty lies in keeping the gauginos and higgsinos 
lighter than $m_{3/2}$.\footnote{To achieve this, it is usually assumed that gravity 
mediation of gaugino masses can be avoided first of all. Furthermore, one has to find ways to suppress 
contributions from anomaly mediation \cite{Randall:1998uk}. Since the latter is not fully understood within 
string theory (see \cite{Antoniadis:2004qn}) and we include the effects of gravity mediation anyway, we will 
not consider anomaly mediation in the following.}
Instead, we propose extra contributions to the masses of charged scalars, 
which make them heavier than $m_{3/2}$, while the fermions including the gravitino remain light.  
\\

The purpose of this paper is to study the confluence of this combined approach with 
F- and D-terms in supergravity and string theory models. By this we mean that we 
assume that some hidden sector dynamics generates F- and D-terms at some scale 
of supersymmetry breaking, but we do not present a full dynamical model, how this happens. 
Instead, we analyze the various scenarios that can emerge in the visible sector, and in particular 
identify classes of models which generically lead to hierarchically split mass scales. 


\subsection{FI-terms in global supersymmetry} 

In many models of grand unification, compactification of higher dimensional supergravity 
or string theory, the minimal gauge symmetries that can be achieved at low energies 
involve various extra abelian $U(1)$ gauge factors beyond the Standard Model gauge group, 
i.e.\ the total gauge symmetry is $SU(3)_C\times SU(2)_L\times U(1)^n$, where 
among the $U(1)$ there is also the hypercharge. In string theory it often happens 
that some of the extra factors are actually anomalous, the anomaly being canceled by a 
(generalized) Green-Schwarz (GS) mechanism, in which the gauge boson develops a 
Stueckelberg mass and decouples (see e.g.\ \cite{Klein:1999im}). In any case, it is then permissible to add 
FI-terms $\xi_a D_a$ to the supersymmetric Lagrangian, one 
for each $U(1)_a$. The D-term potential in global supersymmetry is 
\beqn
\cv_D ~=~ \sum_a \frac{g_a^2}{2} D_a^2  = \sum_a \frac{g_a^2}{2} 
 \Big( \sum_i Q_a^i |\tilde f_i|^2 + \xi_a \Big)^2 
\eeqn
and thus $\xi_a>0$ leads to the formation of a condensate for at least some 
field $\tilde f_i$ of negative charge $Q_a^i<0$. This breaks the gauge symmetry spontaneously, but 
supersymmetry can be restored at the minimum if $\langle D_a \rangle =0$.\footnote{In string theory, 
the FI-parameter is usually a function of the moduli, $\xi_a=\xi_a(T_I, \bar T_{\bar I})$. 
Therefore, turning on the FI-term can correspond to a flat direction $|\tilde f_i|^2 = \xi_a$
in the total potential, for $Q_a^i=-1$.}
In the MSSM it is usually assumed that the FI-term of the hypercharge is 
absent or very small, and does not play a role in the Higgs potential.  \\ 

Whenever the auxiliary field obtains a non-vanishing expectation value $\langle D_a \rangle \not=0$, 
supersymmetry is broken, and mass terms are generated for all the charged scalars, 
\beqn 
m_i^2 ~=~ \sum_a g_a^2 Q_a^i \langle D_a \rangle \ , 
\eeqn 
where it is now assumed that the charges are positive for  the MSSM fields, to avoid 
breaking of the Standard Model gauge symmetries. This scenario can be achieved in a global 
supersymmetric model with a single extra $U(1)_X$ 
by adding two scalars $\phi^\pm$ to the MSSM, singlets under 
$SU(3)_c\times SU(2)_L\times U(1)_Y$, but with charges $\pm 1$ under 
$U(1)_X$ \cite{Dvali:1996rj}. The crucial ingredient is an interaction in the superpotential 
of the form 
\beqn \label{supo}
W_\pm = m \phi^+\phi^-\ . 
\eeqn 
Minimizing the full potential 
\beqn
\cv = m^2 \left( |\phi^+|^2 + |\phi^-|^2 \right) + 
\frac{g_X^2}{2} \Big( \sum_i Q_X^i |\tilde f_i|^2 +|\phi^+|^2-|\phi^-|^2 +\xi_X \Big)^2 
\eeqn
drives the fields to 
\beqn
\langle \phi^+ \rangle =0\ , \quad 
\langle \phi^- \rangle^2 = \xi_X- \frac{m^2}{g_X^2} \ , 
\eeqn
and 
\beqn \label{F-terms1}
\langle D_X \rangle ~=~ \frac{m^2}{g_X^2}\ , \quad 
\langle F_{\phi^+} \rangle ~=~ m \sqrt{\xi_X} +\, \cdots \ . 
\eeqn 
Gaugino masses may originate from higher dimensional operators, and 
are suppressed by powers of 
$M_{\rm Pl}$\footnote{The Planck mass $M_{\rm Pl}$ 
is defined so that $M_{\rm Pl}=\kappa^{-1}=(8\pi G)^{1/2}=2.4\times 10^{18}$ GeV.}, 
\beqn \label{gaugino}
m_\lambda ~\sim~ \frac{1}{M_{\rm Pl}^2} \langle F_{\phi^+}\phi^- + F_{\phi^-}\phi^+ \rangle
          ~\sim~ m \frac{\xi_X}{M_{\rm Pl}^2}\ . 
\eeqn 
Assuming $m \sim \co({\rm TeV})$ and $\xi \sim \co(M_{\rm Pl}^2)$ 
on gets masses at the electro-weak scale. 
Depending on the precise scale and the 
charges of the MSSM scalars under the extra $U(1)_X$, these contributions to 
their masses can be very important in the soft breaking Lagrangian. 
A central point to notice here is the fact that the masses that follow from 
the FI-terms are directly proportional to the expectation values of the auxiliary $D_a$ fields, 
they are mediated by the anomalous $U(1)_X$, whereas the masses induced via the F-terms are 
suppressed by $M_{\rm Pl}$ through their mediation by gravity. The function of 
the extra fields $\phi^\pm$ lies in absorbing the potentially large FI parameter 
$\xi_a$, such that $\langle D_a \rangle^{1/2} \sim \co({\rm TeV})$, consistent with the 
standard scenario of superpartner masses at the electro-weak scale. 
\\

We will discuss this type of model and its modifications in the frame work 
of supergravity and string theory. However, before getting into the details of the 
extended model, we discuss how such FI-terms arise in string theory. 


\subsection{A single anomalous $U(1)$ and the heterotic string} 

The four-dimensional GS mechanism consists of the cancellation 
of the anomalies of the usual one-loop triangle diagrams with tree-level exchange of an axionic 
scalar $\sigma$. This refers to the mixed abelian-gravitational 
and abelian-non-abelian anomalies at the same time. 
The relevant terms in the action are usually written 
in terms of the Hodge-dual 2-form $B_{\mu\nu}$, related to $\sigma$ by 
$\partial_\mu \sigma \sim \epsilon_{\mu\nu\rho\kappa} \partial^\nu B^{\rho\kappa}$, as 
\beqn \label{gscpl}
c_A \partial_\mu B_{\nu\rho} \omega_3^{\mu\nu\rho} + m_X \epsilon^{\mu\nu\rho\kappa} B_{\mu\nu} F^X_{\rho\kappa} 
\eeqn 
where $c_A$ and $m_X$ are two coupling constants,  $\omega_3$ the Chern-Simons 
(CS) 3-form, and $F^X$ the gauge field strength of the relevant $U(1)_X$. 
Now, $\sigma$ is the imaginary part of some complex scalar in a chiral multiplet. 
For the heterotic string, the only such scalar that 
participates in the GS mechanism is the dilaton-axion field $S|_{\theta=\bar\theta=0}=s + i\sigma$ (see 
\cite{Lalak:1999bk} for an overview). 
Its action is described by the K\"ahler potential 
$\ck (S,\bar S)= -\ln(S+\bar S)$. Since Eq.(\ref{gscpl}) implies a non-linear gauge transformation 
$\delta_X S \sim m_X\epsilon_X$ under the $U(1)_X$, the gauge invariance demands a redefinition of the 
K\"ahler potential $\ln(S+\bar S) \rightarrow \ln(S+\bar S - m_X V_X)$, where 
$V_X$ is the vector multiplet superfield. The Lagrangian then involves a Stueckelberg mass term 
with mass proportional to $m_X$ for the gauge boson of this $U(1)_X$, 
which absorbs the scalar $\sigma$ as its longitudinal component. 
In addition, an FI-term $\xi_X D_X$ is present, with $\kappa^2 \xi_X \sim m_X/s$. 
For the heterotic string, this FI-term is generated at one-loop 
and the coefficient reads \cite{Dine:1987xk} 
\beqn 
\kappa^2 \xi_X ~\sim~ \frac{m_X}{s} ~\sim~ \frac{g_X^2 \tr(Q_X)}{192\pi^2} \ .  
\eeqn 
In the presence of an interaction (\ref{supo}) 
the scalar fields $\tilde f_i$ charged under the $U(1)_X$ acquire masses given by
\beqn
m_{i}^2 ~=~ Q_X^i m^2 ~\sim~ \co({\rm TeV})\ . 
\label{mass1}
\eeqn
The remarkable feature of Eq.(\ref{mass1}) is that it is independent of 
the FI-parameter. Thus, the vector boson gets a mass of the order of $m_X$, which is close to 
the Planck scale, whereas the charged sfermions and gauginos 
remain massless at the high scale, and get masses of the order of the electro-weak scale. 
This is the standard scenario of supersymmetry breaking via an anomalous $U(1)$ with GS 
mechanism. 
\\


\subsection{Multiple Anomalous $U(1)$ Symmetries and D-branes} 

Orientifold string compactifications \cite{Angelantonj:2002ct} usually involve more than one 
anomalous $U(1)$ factor. While in the heterotic string it is only the 
axionic partner of the dilaton 
that participates in the GS anomaly cancellation, now 
all the axionic scalars that follow from the reduction of 
the RR forms from ten dimensions can do so \cite{Ibanez:1998qp}. 
In orientifold compactifications of type IIB strings, the 
relevant RR scalars originate from the twisted sectors. 
The FI-parameters $\xi_a$ are then functions of 
the expectation values of the real parts of these twisted scalars, instead of the dilaton $s$. 
For a special example of this class of models, in a toroidal 
orbifold $\mathbb T^6/\mathbb Z_3$, it was 
shown, that no FI-term was generated at one-loop, 
consistent with the fact that the twisted scalars vanish at the 
orbifold point \cite{Poppitz:1998dj}. 
As another class of models, orientifolds with intersecting 
(type IIA) or magnetized (type IIB) D-branes have been studied extensively 
in the recent past, most prominently for their very attractive features to produce 
Standard Model or MSSM like gauge groups and spectra. For these, 
untwisted RR scalars participate in the GS mechanism. Again the FI-term at 
tree-level (i.e.\ from a disc diagram, or the 
dimensional reduction of the Born-Infeld action) is 
proportional to the modulus that combines with the 
axionic scalar from the GS mechanism into a complex scalar. 
The GS couplings analogous to Eq.(\ref{gscpl}) now involve many scalars 
\beqn \label{gs2}
\sum_{I,A} c^I_A \partial_\mu B^I_{\nu\rho} \omega_3^{\mu\nu\rho} 
 + \sum_{a,I} m_a^I \epsilon^{\mu\nu\rho\kappa} B^I_{\mu\nu} F^a_{\rho\kappa} \ , 
\eeqn 
where $I$ labels the scalars $\sigma_I$, given by 
$\partial_\mu \sigma_I \sim \epsilon_{\mu\nu\rho\kappa} \partial^\nu B^{I\rho\kappa}$, and 
$a$ the anomalous $U(1)_a$ factors with field strengths $F^a$ for the superfield $V_a$. 
The constants $c_A^I$ are labeled by $A$ for the different anomalies, i.e.\ the different CS 
forms that can appear. We let 
$T_I|_{\theta=\bar\theta=0} = t_I + i\sigma_I$ be the complex scalars. 
Again Eq.(\ref{gs2}) implies that the $T_I$ transform under $U(1)_a$ whenever the coupling 
coefficient $m_a^I\not=0$. Then the K\"ahler coordinate $T_I$ is replaced in the 
following way 
\beqn 
\ck(T_I+\bar T_I) ~\rightarrow~ \ck(T_I+\bar T_I - \sum_a m_a^I V_a ) \ . 
\eeqn 
Depending on the precise form of the K\"ahler potential, a FI-term will be generated 
from this expression, that will depend on the vacuum expectation value of $t_I$. 
The simplest expression would be  
\beqn 
\kappa^2 \xi_a ~\sim~ \sum_I m_a^I t_I \ . 
\label{13}
\eeqn 
It was stressed in \cite{Ibanez:1998qp} that the $\xi_a$ given by Eq.(\ref{13}) 
can in principle be of any size, as opposed to the result for the heterotic string case.
Another important observation is to note, that the FI-terms are not necessarily tied to 
anomalous gauge symmetries, but only the non-vanishing Stueckelberg coupling $m_a^I\not=0$ 
has to exist.\footnote{As mentioned earlier, we shall here not attempt to model the dynamics of the 
hidden sector in any detail, and therefore also do not try to answer, how 
the FI-parameters are generated dynamically. 
Since they are moduli-dependent functions, a meaningful answer would 
have to address the moduli-stabilization at the same time. 
} 


\section{FI-terms in supergravity and string theory}

We now examine the patterns of soft supersymmetry breaking that arise 
from an effective string theory Lagrangian with one or more FI-terms, motivated by 
the appearance of multiple $U(1)$ factors in orientifold models, that can develop FI-terms. 

\subsection{The vacuum energy} 

The degrees of freedom of the model are assumed to be given by the fields $f_i$ of the MSSM, 
the extra gauge vector multiplets for the $U(1)_a$, the moduli $T_I$ of the gravitational sector, plus 
the axion-dilaton $S$, which includes the fields that participate in the GS mechanism, producing 
Stueckelberg masses for the gauge bosons and FI-terms. Furthermore, we can add extra fields like the 
$\phi^\pm$ of the globally supersymmetric model, 
with charges $\pm \alpha_a$ under $U(1)_a$. The effective scalar potential
is  given by the $\cn=1$ supergravity formula \cite{Cremmer:1982en}
\beqn
\cv ~=~ - \kappa^{-4}e^{-G} [G^{M\bar N}G_MG_{\bar N} +3] + {\cal V}_D \ , 
\eeqn 
with 
\beqn 
G ~=~ - \kappa^2  \ck- \ln(\kappa^6 WW^{\dagger}) \ , 
\eeqn
where indices $M,N$ run over all fields. 
We define the dilaton and moduli fractions of the vacuum energy by 
\beqn
|\gamma_S|^2=-\frac{1}{3} G^{S\bar S}G_SG_{\bar S}\ , \quad 
|\gamma_I|^2=-\frac{1}{3}
G^{I\bar I}G_IG_{\bar I}\ , 
\eeqn
and $|\gamma_{\pm}|^2$ in a similar fashion. 
It also turns out to be useful to introduce the following combinations  
\beqn
&& 
m_{3/2} = \kappa^{-1} e^{-G/2}\ , \quad 
m_\pm = e^{\kappa^2 \ck/2} m \ , \quad  
x = \kappa \langle \phi^+ \rangle \ , \quad  
y = \kappa \langle \phi^-\rangle \ . \quad  
\eeqn
Similarly, all other fields are made dimensionless. 
Imposing the restriction on the model  
that the vacuum energy vanishes (through fine tuning) one has 
\beqn
|\gamma_S|^2 + \sum_I |\gamma_I|^2 + |\gamma_+|^2+ |\gamma_-|^2
+\frac{1}{3m_{{3}/{2}}^2M_{\rm Pl}^2} 
 \sum_a \frac{g_a^2}{2} D_a^2  =1\ , 
\eeqn
where $D_a =  \alpha_a |\phi^+|^2-\alpha_a |\phi^-|^2 + \xi_a$. 
This implies an immediate bound on the expectation values of the auxiliary fields
$F_I = D_I W= \partial_I W + \kappa^2(\partial_I \ck) W$ and $D_a$, 
\beqn 
\langle F_I \rangle ~\lesssim~ m_{{3}/{2}} M_{\rm Pl}\ , \quad 
\langle D_a \rangle ~\lesssim~
m_{{3}/{2}} M_{\rm Pl}\ ,  
\label{19}
\eeqn 
where we ignore prefactors involving the K\"ahler potential. 
As long as $m_{{3}/{2}}\sim\co({\rm TeV})$, Eq.(\ref{19}) implies 
roughly $\langle D_a \rangle^{1/2} \lesssim 10^{10-13}\, {\rm GeV}$, 
which is the usual intermediate supersymmetry breaking scale in SUGRA models. 
The masses that are generated by the F-terms are given by  
$F_I/M_{\rm Pl}\sim\co({\rm TeV})$, whereas the D-terms would be 
able to produce much larger mass terms proportional to $D_a^{1/2}$. 
This means that the mass parameter in the superpotential (\ref{supo}), which 
had to be fine tuned to the electro-weak scale, can now also be assumed as large as the 
intermediate scale, $m\lesssim \sqrt{m_{{3}/{2}}M_{\rm Pl}}$. 
Further, we note that the scenario with a Planck 
scale sized FI-parameter, as is unavoidable for the 
heterotic string in the presence of an anomalous $U(1)$, 
is only consistent with a Planck scale sized gravitino mass. 
In orientifold D-brane models, as mentioned earlier, the FI-parameter can in principle 
be of any value, and the problem does not occur. 
\\

In scenarios with split supersymmetry, the gravitino mass itself is not restricted to a small value. However, 
gravity mediation generically leads to a contribution to gaugino and higgsino masses which 
is proportional to the gravitino mass, and therefore $m_{{3}/{2}}\sim\co({\rm TeV})$ 
is unavoidable in the present context. This then really puts an upper bound $10^{10-13}\, {\rm GeV}$ on 
the high mass scale allowed for the sleptons and squarks. 
\\

Before going into the various scenarios, let us first assemble a few general definitions and formulas 
for the supergravity version of the model of supersymmetry breaking mediated by one or many anomalous $U(1)$.  
For the K\"ahler potential $\ck$ we write 
\beqn
\ck &=& \ck_{\rm hid}(T_I,\bar T_{\bar I}) + 
\ck_{i\bar\imath}(T_I,\bar T_{\bar I}) f_{i}f_{\bar\imath}^{\dagger} 
+ \ck_+(T_I,\bar T_{\bar I})  \phi^+\phi^{+\dagger} 
 + \ck_-(T_I,\bar T_{\bar I})  \phi^-\phi^{-\dagger} \ ,  
\nonumber 
\eeqn
where we have now included $S$ among the $T_I$. 
The gauge kinetic functions are moduli-dependent, 
\beqn 
f_a ~=~ f_a(T_I)\ , \quad \frac{1}{g_a^2}~=~ \Re(f_a) \ , 
\eeqn 
but independent of $\phi^\pm$. 
For the superpotential we assume the following factorized form
\beqn
W ~=~ W_{\rm MSSM}(f_i) + W_{\rm hid}(T_I) + W_\pm(\phi^+,\phi^-) ~=~ m\phi^+\phi^- + W_0 
\eeqn
where $W_{\rm MSSM}$  contains the quark, lepton and Higgs fields, $W_{\rm hid}$
contains the fields of the hidden sector which break supersymmetry spontaneously  
by generating auxiliary field components for 
$F_I = D_IW_{\rm hid}$, while $W_\pm$ is still given by 
Eq. (\ref{supo}).\footnote{This implies an assumption on the absence of any coupling among MSSM fields 
and $\phi^\pm$ in the superpotential, which may be problematic in the context of a concrete model.
Furthermore, we also ignored any cross-coupling in the K\"ahler potential, where in principle the 
moduli-dependence of the various coefficients could also involve $\phi^\pm$.}  
With this, the total D-term potential ${\cal V}_D$ is given by
\beqn
{\cal V}_D =  {\cal V}_D^{\rm MSSM} + \sum_a\frac{g_a^2}{2} \Big(\sum_i Q_a^i \ck_{i\bar\imath} |\tilde f_i|^2  
+\alpha_a \ck_+ |\phi^+|^2-\alpha_a \ck_- |\phi^-|^2 +\xi_a\Big)^2\ . 
\label{d1}
\eeqn 
Here ${\cal V}_D^{\rm MSSM}$ is the D-term arising from 
the $SU(2)_L\times U(1)_Y$ sector, which will not be important.
The standard expressions for the soft breaking terms that originate from the F-terms only, 
are \cite{Kaplunovsky:1993rd}
\beqn 
m_\lambda ~=~ \frac{1}{2\Re(f_a)} F^I \partial_I f_a\ , 
\eeqn 
for gaugino masses, and 
\beqn 
m_{{\rm gr,}i}^2 ~=~ m_{3/2}^2 - F^I \bar F^{\bar J} \partial_I \partial_{\bar J} \ln( \ck_{i\bar\imath} ) 
\eeqn 
for scalar masses. For $F_I \sim m_{3/2}M_{\rm Pl}$, both masses are of the order of $m_{3/2}$. 


\subsection{The simplest model}

The simplest model that already displays the effects of the FI-terms is given 
by assuming one or more FI-terms being generated by extra $U(1)_a$ gauge factors, and only 
including the MSSM fields with arbitrary positive charges, but leaving out the 
extra fields $\phi^\pm$.\footnote{It may sound very restrictive to allow only positive charges here, 
and it really would be in any reasonable model derived from a GUT or string theory. However, 
there are well known cases in string theory compactifications, where higher order corrections 
in the derivative expansion of the effective action (such as the Born-Infeld Lagrangian) 
lead to a lifting of tachyonic negative masses in the presence of FI-terms, even if some fields have 
negative charge. We will come to explain this in some more detail later.}
In that case, supersymmetry is broken, and the D-terms are trivially given by 
\beqn 
\frac{g_a^2}{2} \langle D_a^2 \rangle ~=~ \frac{g_a^2}{2} \xi_a^2\ . 
\eeqn 
Together with potential F-terms, they generate masses 
\beqn 
m_i^2 ~=~ m_{{\rm gr,}i}^2 + \sum_a g_a^2 Q_a^i \ck_{i\bar\imath} \xi_a
\eeqn 
for all charged scalars. 
On the other hand, the masses of gauginos (and higgsinos) are unaffected by the D-term, at least at leading 
order, and would be dictated by the F-terms to be of order $m_{3/2}$. 
In a scenario with $m_{3/2}\sim \co({\rm TeV})$, and $D_a \sim 10^{10-13}\, {\rm GeV}$ for at least one 
FI-term, this provides a hierarchical split of energy scales, however, the high scale cannot move 
up all the way to the Planck scale. The most simple charge 
assignment would give positive charges of order one to all MSSM fields, except the Higgs, 
and thus the sfermion sector 
of the MSSM would become very heavy and undetectable at LHC.\footnote{This charge assignment would 
indeed render the $U(1)$ anomalous.} 
There may of course also be other interesting 
patterns to consider, such as different charges for the three generations, different charges for 
different $SU(5)$ multiplets, which would leave the gauge coupling unification intact, 
or different charges for left- and right-handed fields, which could be more easily realized in 
certain D-brane models from string theory. 


\subsection{The full potential with a single $U(1)_X$}

The essence of the model of \cite{Dvali:1996rj}, 
where an anomalous $U(1)$ with its FI-term is responsible for 
supersymmetry breaking, is the scalar condensate for $\phi^-$ that cancels the contribution 
of the FI-parameter to the D-term, up to small electro-weak scale sized contribution, given the 
interaction (\ref{supo}) in the superpotential. Since such a condensate breaks the gauge symmetry, 
one has to add extra chiral multiplets $\phi^\pm$ to the MSSM, neutral under the MSSM gauge 
symmetries. The model of \cite{Dvali:1996rj} is within the framework of global supersymmetry, and it
 is of obvious interest to study the embedding of this original model into supergravity. 
So, we focus first on the case when there is just one extra $U(1)_X$, 
which develops a FI-term. 
The potential can be written as 
\beqn
\kappa^4 \cv &=& \kappa^4 \cv_{\rm hid} + 
e^{\kappa^2 \ck} \Big[  \kappa^2 m^2 \left( |x|^2 + |y|^2 \right) 
 + \kappa^6 |W|^2 \left( |\ck_+ x|^2 + |\ck_- y|^2 -3 \right)
 \\ 
&& 
\hspace{.5cm} 
+~ \kappa^4 m \left( \ck_+ + \ck_- \right) \left( xy \bar W+\bar x\bar y W\right) \Big] 
+ \frac{g_X^2}{2}  \Big [ \ck_+|x|^2 - \ck_-|y|^2 +\xi_X \Big]^2 \ , 
\nonumber 
\eeqn 
where we also have replaced $\xi_X \rightarrow \kappa^{-2} \xi_X$, and defined 
\beqn 
 \cv_{\rm hid}   ~=~  e^{\kappa^2 \ck} \ck^{I\bar J} D_I W D_{\bar J}\bar W 
\eeqn 
for the contribution of the hidden sector. It is essential for the fine tuning of the vacuum energy. 
The two minimization conditions are $\partial_x V = \partial_y V = 0$. To keep things simple, 
we now also  set all relevant phases to zero, i.e.\ we treat $x,y,W$ as real. 
Then one gets the following two equations
\beqn
&& \ck_+ x [ \kappa^2 g_X^2 D_X + \kappa^4  \cv_F] + \frac{m_\pm^2}{M_{\rm Pl}^2} [x+ xy^2 (\ck_+ + \ck_-)] 
 \\
&& \hspace{3cm} + \frac{m_{3/2}^2}{M_{\rm Pl}^2} \ck_+^2 x
  + \frac{m_\pm m_{3/2}}{M_{\rm Pl}^2} y [(\ck_+ x)^2+(\ck_- y)^2 +2\ck_- -3] ~=~0\ , 
\non
&&
\ck_- y [- \kappa^2 g_X^2 D_X + \kappa^4  \cv_F] + \frac{m_\pm^2}{M_{\rm Pl}^2} [y+ x^2y (\ck_+ + \ck_-)] 
 \non
&& \hspace{3cm} + \frac{m_{3/2}^2}{M_{\rm Pl}^2} \ck_-^2 y
  + \frac{m_\pm m_{3/2}}{M_{\rm Pl}^2} x [(\ck_+ x)^2+(\ck_- y)^2 +2\ck_+ -3] ~=~0\ . 
\nonumber 
\eeqn 
where  $\cv_F$ and $D_X$ are defined so that $\cv = \cv_F + \frac12 g_X^2 D_X^2$,
where $\cv_F$ and $\cv_{\rm hid}$ are related by 
\beqn
\cv_F &=&  \cv_{\rm hid} + 
 \Big[   m_{\pm}^2 M_{\rm Pl}^2 \left( |x|^2 + |y|^2 \right) 
 +  |m_{3/2}|^2 M_{\rm Pl}^2\left( |\ck_+ x|^2 + |\ck_- y|^2 -3 \right)
 \\ 
&& 
\hspace{1cm} 
+~  m_{\pm}M_{\rm Pl}^2 \left( \ck_+ + \ck_- \right) \left( xy \bar m_{3/2}+\bar x\bar y m_{3/2}\right) \Big], 
\nonumber 
\eeqn 
Note that the redefined mass parameters are field-dependent. 
In the following, we also restrict to canonically normalized fields, and set 
$\ck_+=\ck_-=1$.  
We have illustrated the full potential in figure 1, using values 
\beqn \ 
\kappa^4 \cv_{\rm hid} ~=~ 3\ , ~~
\kappa^6 e^{\kappa^2\ck}|W_0|^2 =1\ , ~~   
\kappa^2 e^{\kappa^2\ck} m^2=1 \ , ~~ 
g_X=\sqrt{200} \  , ~~
\kappa^2\xi_X=10^{-1}\ . 
\nonumber
\eeqn 
This corresponds to setting some parameters equal, dividing the total potential 
by $M_{\rm Pl}^2 m_{3/2}^2$, and rescaling the Planck mass by many order of magnitude,  
so that $M_{\rm Pl}/m_{3/2}=10$, just 
to suppress the very steep part of the potential. It turns out, that for the relevant 
range of parameters $x\neq 0 \neq y$ at the minimum. 

\begin{figure}[h]
\begin{center}
  \resizebox{6cm}{!}{\psfig{figure=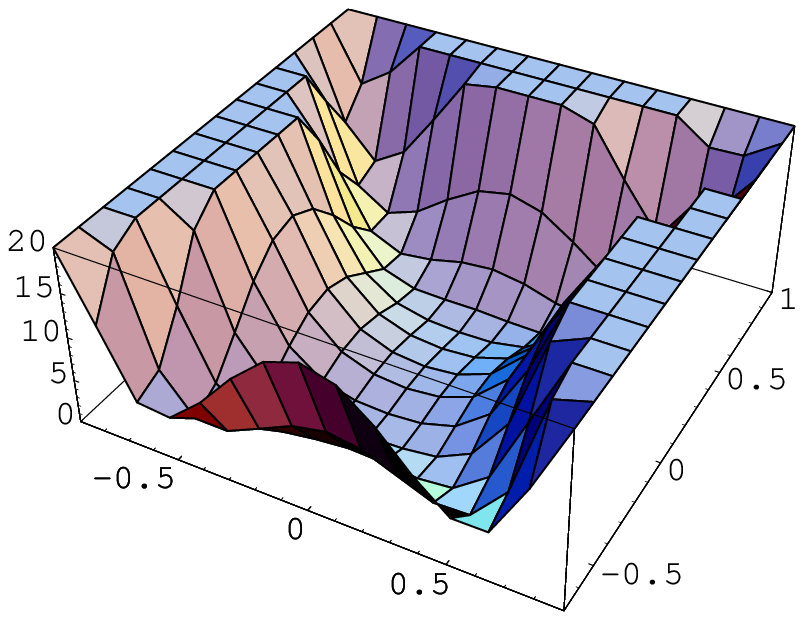,width=4cm}}
\caption{Rescaled potential $\cv(x,y)$} \label{pot1}
\end{center}
\end{figure}
Since the minimization of the F-term potential basically consists of balancing 
terms that scale like $m_\pm^2 |\phi^\pm|^2$, $m_{3/2}^2 |\phi^\pm|^2$ 
or with inverse powers of $M_{\rm Pl}$, with the negative term 
$m_{3/2}^2 M_{\rm Pl}^2$, it is intuitively clear that $x$ and $y$ will 
be roughly bounded through the most dangerous term by
$m_{3/2} m^{-1}_\pm$ or $\co(1)$. This is of course only 
valid as long as $|\phi^\pm|^2 \gg \xi_X$, otherwise $|\phi^-|^2$ gets tied to $\xi_X$. 
In any case one has that $F_{\phi^\pm}/M_{\rm Pl}\lesssim m_{3/2}$, which is sufficient for light gauginos. 


\subsection{Limiting cases: $m_{3/2}=0$ or $m_\pm=0$}

We consider now some limiting cases for the above. To see how the 
model of \cite{Dvali:1996rj} emerges, we 
take the flat limit by setting 
\beqn
m_{3/2}~=~ 0\ . 
\eeqn
The two minimization conditions reduce to equations homogeneous in $x,y$ respectively. 
One can convince oneself that $x=0$ is stable. The 
solution then reduces to the known case of global supersymmetry, where 
\beqn\label{sol1}
x~=~0\ , \quad y^2 ~=~ \xi_X + 1 \pm \sqrt{ 1+\frac{2m_\pm^2}{g_X^2M_{\rm Pl}^2}} 
                 ~\sim~ \xi_X - \frac{m_\pm^2}{g_X^2M_{\rm Pl}^2} \ . 
\label{m32zero} 
\eeqn
The auxiliary fields are also identical to (\ref{F-terms1}), 
since $D_I W =\partial_I W$. Thus, in this limit, all masses are of the order of $m_\pm\sim m$. 
\\

On the other hand, it is interesting to study the supergravity corrections to the case that 
allowed to restore supersymmetry in global supersymmetry, when the 
mass term in the superpotential is absent, 
\beqn 
m_\pm ~=~ 0\ . 
\eeqn
In this case the solution is given by 
\beqn 
x^2 + y^2 ~=~ 2 - \frac{\cv_{\rm hid}}{m_{3/2}^2 M_{\rm Pl}^2} \ , \quad 
x^2 - y^2 ~=~ - \xi_X \ .
\label{mzero}
\eeqn 
Obviously, there is no supersymmetry breaking by D-terms, and $\langle D_X\rangle =0$. 
Actually, in this limit the vacuum energy 
cannot be fine tuned without extra contributions to the potential, irrespective of $\cv_{\rm hid}$, 
since it is always given by $- m_{3/2}^2M_{\rm Pl}^2$. This is however an artifact of the limit $m_\pm=0$, and 
for non-vanishing $m_\pm$, the value at the minimum 
can be shifted by varying $\cv_{\rm hid}$. Finally as a special case of 
Eq.(\ref{mzero}), one may consider the case 
\beqn
\cv_{\rm hid}~=~ (2-\xi_X)m_{3/2}^2M_{\rm Pl}^2
\eeqn
which gives 
\beqn
x~=~0\ , \quad y^2~=~\xi_X\ . 
\label{mzero1}
\eeqn
The values of Eq.(\ref{mzero1}) give  the minimum of the potential that is in some sense 
analogous to the case Eq.(\ref{m32zero}).


\subsection{Multiple $U(1)$ gauge symmetries} 

With multiple FI-terms in the potential, it is clear that supersymmetry breaking 
can occur more generically. We have seen above, that FI-terms that cannot be canceled by 
scalar condensates lead to large scale masses, whereas a scalar condensate with 
a single FI-term was able to lower the masses to the electro-weak 
scale $m_{3/2}$, similar to \cite{Dvali:1996rj}.  
If multiple FI-terms now come accompanied by the same number of extra charged fields to 
develop condensates, then the Lagrangian would just be a sum of identical copies of the one 
of the previous sections, and nothing new is found. An interesting case arises, when 
there is a mismatch, and not all the FI-parameters can be absorbed by fields like $\phi^-$, 
and thus some large masses can be generated. 
We now analyze the situation, when there still is only a single set of extra charged 
fields $\phi^\pm$, that takes a non-vanishing expectation value, but there are multiple $U(1)_a$ 
gauge symmetries, under which it is charged.  
The potential is only modified by summing over D-terms, 
\beqn
\kappa^4 \cv_D &=& 
\sum_a \frac{g_a^2}{2}  \Big [ \alpha_a \ck_+ |x|^2 - \alpha_a\ck_- |y|^2 +\xi_a \Big]^2 \ , 
\nonumber 
\eeqn 
again making $\xi_a$ dimensionless by $\xi_a\rightarrow\kappa^{-2} \xi_a$. 
Regarding the vacuum energy cancellation, the new minimization conditions read
\beqn
&& \ck_+ x \Big[ \kappa^2 \sum_a \alpha_a g_a^2 D_a + \kappa^4  \cv_F \Big] 
 + \frac{m_\pm^2}{M_{\rm Pl}^2} [x+ xy^2 (\ck_+ + \ck_-)] 
 \\
&& \hspace{3cm} + \frac{m_{3/2}^2}{M_{\rm Pl}^2} \ck_+^2 x
  + \frac{m_\pm m_{3/2}}{M_{\rm Pl}^2} y [(\ck_+ x)^2+(\ck_- y)^2 +2\ck_- -3] ~=~0\ , 
\non
&&
\ck_- y \Big[ -\kappa^2 \sum_a \alpha_a g_a^2 D_a + \kappa^4  \cv_F \Big]  
 + \frac{m_\pm^2}{M_{\rm Pl}^2} [y+ x^2y (\ck_+ + \ck_-)] 
 \non
&& \hspace{3cm} + \frac{m_{3/2}^2}{M_{\rm Pl}^2} \ck_-^2 y
  + \frac{m_\pm m_{3/2}}{M_{\rm Pl}^2} x [(\ck_+ x)^2+(\ck_- y)^2 +2\ck_+ -3] ~=~0\ . 
\nonumber 
\eeqn 
It is not necessary to study the general solution here, since the multiple FI-terms already 
break supersymmetry without the superpotential (\ref{supo}), and one can therefore set $m_\pm=0$. 
With $\ck_+=\ck_-=1$ one can easily solve for $x,y$, finding  
\beqn 
x^2 + y^2  ~=~ 2 - \frac{\cv_{\rm hid}}{m_{3/2}^2 M_{\rm Pl}^2} \ , \quad 
x^2 - y^2  ~=~ - \frac{\sum_a g_a^2 \xi_a }{M_{\rm Pl}^2 \sum_a g_a^2}\ . 
\eeqn 
Thus, in a generic situation, where not all $\xi_a$ are equal, supersymmetry will be 
broken, and $\langle D_a\rangle \sim \xi_a \sim m_{3/2} M_{\rm Pl}$. 
In this case, the vacuum energy is given by the negative F-term contribution plus D-terms, and can 
be fine tuned even without $\cv_{\rm hid}$. 
The F-terms are given by $F_{\phi^\pm}=\kappa^2 \phi^\pm W_0 \lesssim m_{3/2}M_{\rm Pl}$. 
The contributions to the scalar masses now look like (with dimensionful $\xi_a$) 
\beqn 
m_i^2 ~=~ m_{{\rm gr},i}^2 + \sum_a g_a^2 Q_a^i D_a ~\sim~ \sum_a g_a^2 Q_a^i \xi_a +\, \cdots
\eeqn 
where the terms in parentheses can be of the same order of magnitude, but the 
gravity-mediated contributions are negligible. This realizes the split supersymmetry scenario, 
if $\xi_a\sim 10^{10-13}\, {\rm GeV}$, for at least two FI-parameters, but with 
a low gravitino mass $m_{3/2}\sim \co({\rm TeV})$. 
\\

It may also be interesting to note that at the same time all other soft breaking 
parameters will get modified in the presence of FI-terms and scalar field condensates. 
This happens through the prefactor $e^{\kappa^2\ck}$ in the total potential. 
For instance, the bi-linear couplings $B$ and the tri-linear couplings $A$ are (see e.g.\ \cite{Nath:2002nb}) 
\beqn 
A_{\alpha\beta\gamma}^0\ , ~~B_{\alpha\beta}^0 ~\propto~
m_{3/2} e^{\kappa^2\ck/2} ~=~ m_{3/2} (S+\bar S)^{1/2} e^{\frac12 (|x|^2 + |y|^2) + \, \cdots} \ , 
\eeqn
which may lead to extra suppression factors, depending on the model. Further, we note that the 
Higgs mixing parameter $\mu$ that enters in the superpotential in the form $\mu H_1H_2$, 
where $H_1$ and $H_2$ are the two Higgs doublets of MSSM, remains essentially unaffected. 
This is so because
in string/supergravity scenarios one expects the $\mu$ term to arise from the K\"ahler 
potential using a 
K\"ahler tranformation after the spontaneous breaking of supersymmetry has taken place. The K\"ahler
transformation is sensitive to the F-term breaking and not the D-term. Thus, one expects 
the same mechanism that produces a $\mu$ term of electro-weak size for supergravity models to 
hold in this case as well. 


\subsection{Scenarios with partial mass hierarchies}

To summarize, there are several scenarios that are possible, 
which would lead to different patterns in the mass spectrum:
$(i)$ There is only one FI-term, and an extra scalar field $\phi^-$ beyond the 
MSSM fields which forms a vacuum expectation value. Here all the soft scalar masses will be of 
electro-weak size. $(ii)$ In the case of two or more $U(1)_a$ with non-vanishing $\xi_a$, 
and the charges $Q^i_a$ non-zero and sufficiently generic, the scalar masses are 
of the order of $\xi_a$, while the gauginos stay light. One can thus get 
a hierarchical splitting of scalar and fermion masses. 
$(iii)$ One may also achieve hybrid 
scenarios, when only some of the charges $Q_a^i$ are non-vanishing, or 
such that the high mass terms just cancel out. Below we consider two
specific scenarios with partial splitting of scalars.
The implications of this partially split scenario will be very 
different from the usual SUGRA scenario and also from the high scale
supersymmetry scenario of split supersymmetry.


\subsubsection{Model I: $2+1$ generations} 

As the first model we consider the case with  
non-vanishing charges for the first and second
generation of squarks and sleptons, but vanishing charges for the
third generation. In this circumstance 
the former will
develop super heavy masses and will not be accessible at the LHC, 
as opposed to the third generation. 
The above implies that the dangerous flavor changing neutral currents will be automatically
suppressed. Some of the signals of this scenario will be very unique, such as 
the decay of the gluino ($\tilde g$). In SUGRA its 
decay modes are as follows
\beqn
\tilde g ~\rightarrow~ \bar u_i\tilde u_i\, ,\ \bar d_i \tilde d_i
~\rightarrow~ 
\left\{ { u_i \bar u_i \tilde \chi_j^0\, ,\ d_i\bar d_i \tilde \chi_j^0
 \atop u_i \bar d_i \tilde \chi_k^-\, ,\ d_i\bar u_i \tilde \chi_k^+ } \right. \ . 
\eeqn
where $i=1,2,3$ are  the generational indices, $j=1,2,3,4$, 
and $k=1,2$ for neutralinos and charginos.
However, for the case of the model under consideration
 the decay through the first two generations is highly suppressed, and the gluino 
decay can only occur through third generation squarks 
via the modes $\tilde g\rightarrow b\bar b \tilde \chi_j^0$ with
admixtures of $\tilde g\rightarrow b\bar t \tilde\chi_k^+$, 
$\tilde g\rightarrow t\bar b \tilde\chi_k^-$, and  $\tilde g\rightarrow t\bar t \tilde\chi_j^0$, 
depending on the gluino mass. 
There will be no contribution to the anomalous magnetic moment of the muon at the one-loop level.
The following set of phases can arise: $\theta_{\mu}$, $\xi_i$,
and $\alpha_A$, where $\theta_{\mu}$ is the phase of the Higgs  mixing parameter $\mu$, 
$\xi_i$ ($i=1,2,3$) are the phases of the $SU(3)_c$, $SU(2)_L$ and $U(1)_Y$ gaugino masses, 
and $\alpha_A$ is the phase of the common trilinear coupling for the third generation scalars
(However, it should be kept in mind that not all the phases are independent and only 
certain combinations enter in physical processes).
Because of the super heavy nature of the first two generations, the one-loop 
contributions to the electric dipole mement (edm) of the electron and of the neutron are
highly suppressed. However, there can be higher loop corrections to the edms. Specifically,
the neutron edm can get a contribution from the CP violating dimension six operator. 
Unification of gauge coupling constants at the one-loop level will
remain unchanged, although at the two-loop level there will be effects
from the splittings. In this model the staus can be light and thus 
co-annihilation of the LSP neutralinos with the staus can occur, allowing
for the possibility that the neutralino relic density could be consistent with the current data.
Finally, proton decay from dimension five operators 
would not arise via dressings from the first and second generation 
squarks and sleptons, but can arise from the dressings of the 
third generation sfermions. 


\subsubsection{Model II: ${\bf 5}+{\bf 10}$ split spectrum}

As another illustrative example, we consider a model where the squarks and sleptons belonging to
a {\bf 10} of $SU(5)$ have non-vanishing charges and high scale masses, while the
squarks and sleptons belonging to a {\bf 5} have vanishing
charges. In this case, the only light scalars aside from Higgs 
bosons will be $\tilde d^C_i, \tilde e_{Li}, \tilde \nu_i$ ($i=1,2,3$).
In this model, unlike the case of model I, there is a one-loop supersymmetric 
contribution to the muon anomalous magnetic moment $g_{\mu}-2$. 
The CP phases in this model consist of $\theta_{\mu}$ and $\xi_i$ ($i=1,2,3$). 
There also is a one-loop supersymmetric contribution to the edms of 
the electron and of the neutron. 
Further, the  decays of the gluino, the chargino and the neutralino can 
occur only via a smaller subset of states and thus their decay widths 
will be relatively smaller, though they will still decay within the 
detection chamber. Finally proton decay through dimension five operators
 will be highly suppressed in this model and the dominant decay will
 occur through the usual vector boson interactions.


\section{Hierarchical Breaking in D-brane models} 

We turn now to the question of how the models discussed so far fit into 
string theory, in particular in the class of 
intersecting or magnetized D-brane models \cite{Blumenhagen:2000wh,Ibanez:2001nd}. 
These are models within orientifold string compactifications of type II strings with 
D-branes that wrap parts of the internal compactification space, and either with  
magnetic field backgrounds on the brane world volume (in type IIB) or with the branes intersecting 
non-trivially on the internal space (in type IIA) \cite{Bachas:1995ik}. 
For these models a great deal about the Lagrangian of their low energy field theory 
description is known, e.g.\ in \cite{Cremades:2002te,Kors:2003wf}, 
and the soft supersymmetry breaking terms in the conventional 
setting with F-terms generated in the hidden sector, have been determined. 
Furthermore, it is known how FI-terms can 
appear \cite{Douglas:1996sw}.\footnote{These models have also been recently discussed in 
the context of split supersymmetry in \cite{Antoniadis:2004dt}.} 
\\

The gauge group for any single stack of $N_a$ D-branes is $U(N_a)=SU(N_a)\times U(1)_a$ (only 
sometimes an $Sp(N_a)$ or $SO(2N_a)$ subgroup thereof), and thus involves 
extra abelian $U(1)_a$ factors, when the Standard Model is engineered. 
In the first models that were constructed to reproduce the 
non-supersymmetric Standard Model \cite{Ibanez:2001nd}, 
there are four extra $U(1)_a$, including the anomaly-free 
hyper charge and gauged $B-L$ quantum numbers, but also two extra $U(1)$ factors which 
are anomalous. The relevant anomaly is actually a mixed abelian-non-abelian anomaly, i.e.\ 
there are triangle diagrams of the type $SU(N_b)^2-U(1)_a$ non-vanishing for either 
$SU(2)$ or $SU(3)$. In particular, $\tr(Q_a)=0$, and there is no gravitational anomaly. 
While the first models were non-supersymmetric from the beginning, the structure of charge 
assignments, given in table 1 of \cite{Ibanez:2001nd}, 
and the number of $U(1)_a$ can serve as a representative example. 
From this it is clear that these models at least contain two candidate $U(1)_a$, which may 
develop FI-terms. In generality, it is known that, when D-branes violate supersymmetry, this 
is reflected by FI-terms in the effective theory \cite{Douglas:1996sw}. 
The violation of supersymmetry translates into a 
violation of the $\kappa$-symmetry on their world volume, and is geometrically captured by a violation of the 
so-called calibration conditions. For intersecting D-branes models, this has a very simple 
geometric interpretation. Any single brane wraps a three-dimensional internal space. In 
the case of an orbifold compactification it is characterized by three angles $\varphi_i^a$, $i=1,2,3$, 
one for each $\mathbb T^2$ in $\mathbb T^6=(\mathbb T^2)^3$, measuring the relative 
angle of the D-brane with respect to some reference orientifold plane. The supersymmetry condition 
reads $\varphi_1^a\pm \varphi_2^a\pm \varphi_3^a=0\, {\rm mod}\, 2\pi$, with some choice of signs. 
The FI-parameter at leading order is proportional to the deviation, 
\beqn 
\alpha' \xi_a ~\sim~  \varphi_1^a\pm\varphi_2^a\pm\varphi_3^a\ {\rm mod}\ 2\pi\ . 
\eeqn 
The angles are moduli-dependent quantities, and thus the question if a FI-term is generated 
cannot be finally answered without solving the moduli stabilization problem for 
the relevant moduli. In the mirror symmetric 
type IIB description with magnetized branes this is manifest, and 
the relation reads 
\beqn 
\alpha' \xi_a ~\sim~  \frac{f_a^1}{T_1+\bar T_{\bar 1}}\pm\frac{f_a^2}{T_2+\bar T_{\bar 2}}
 \pm\frac{f_a^3}{T_3+\bar T_{\bar 3}} \ ,  
\eeqn 
where $f_a^i/(T_i+\bar T_{\bar \imath}) = \tan(\varphi_i^a)$, and the $T_i$ are the three 
(dimensionless) moduli, whose real parts measure the sizes of the three $\mathbb T^2$, while the 
$f_a^i$ are rational numbers.  
The natural scale for the FI-parameter is the string 
scale $M_s=(\alpha')^{-1/2}$, and a suppression means that the right-hand-side is small 
numerically. 
\\

Another very important property of the string theoretic embedding of D-terms is the fact 
that tachyonic masses (negative mass squared) can be lifted to positive values. 
Inspecting the charge assignments e.g.\ in  \cite{Ibanez:2001nd}, 
one realizes that it is not feasible to have 
positive charges under the $U(1)_a$ for all the fields of the MSSM. This would naively mean 
that some $m_i^2 \sim g_a^2 Q_a^i D_a$ are negative, which would lead to a breakdown of 
gauge symmetry. However, in the particular case 
of orbifold models, the exact string quantization can be performed, and the mass spectrum 
computed for small FI-parameters, without using effective field theory. 
It turns out that for small angles ($\varphi_i^{ab}=\varphi_i^a-\varphi_i^b<\pi/2$), 
the mass of the lowest excitation of two intersecting branes $a$ and $b$ is given by 
\beqn 
\alpha' m_i^2 = \frac12 \sum_{i=1}^3 |\varphi_i^{ab}| - {{\rm max}_{i}}\{ |\varphi_i^{ab}| \} \ , 
\eeqn 
which, for a proper choice of signs, 
vanishes precisely if $\xi_a=0$, consistent with the effective description. 
However, depending on the angles, there is a region in parameter space, where the 
deviation from $\xi_a=0$ only induces positive squared masses, and 
no tachyons (see e.g.\ \cite{Rabadan:2001mt} in 
this context). This comes as a surprise from the low energy point of view, and is explained by 
the presence of higher order corrections in the derivative expansion of the Born-Infeld effective 
action, which become important for strings stretching between intersecting branes \cite{Hashimoto:2003pu}. 
The conclusion is, that the effective mass that follows from the D-term potential 
is corrected to positive values, and we can tolerate 
tachyons in the field theoretic models. 
\\

Taking these observations together, it seems first of all possible that D-brane models 
of the type discussed have mass spectra with important contributions from FI-terms. Since multiple 
extra $U(1)_a$ factors exist and turn anomalous, a scenario, where the $D_a$ fields cannot be relaxed to the 
electro-weak scale, appears plausible. This would imply a mass hierarchy between charged scalars 
and fermions. To solve the moduli stabilization problem, and show convincingly how 
the D- and F-terms of the desired magnitude are generated dynamically, of course, remains an open challenge. 
One may also want to turn the argument around, to conclude that within the conventional 
approach with low energy supersymmetry, the potential presence of many FI-terms is a great danger 
for these types of models, and one has to find ways to suppress them dynamically. 
\\

Finally, we like to mention a caveat which makes the classes of D-branes models that we 
discussed rather not so good candidates to realize split supersymmetry. It is well known 
that in D-brane models in general the unification of gauge interactions really only happens at the 
string scale, and not in the field theory regime. This means that the original motivation 
to keep the gauginos and higgsinos light, while giving up on the scalars, is upset. 
The gauge kinetic functions 
are moduli-dependent $f_a=f_a(T_I)$, and a unification of couplings requires some 
relations among moduli to hold \cite{Blumenhagen:2003jy,Kors:2003wf,Antoniadis:2004dt}, 
which so far can appear accidentally in various models 
in the literature, but do not seem to have any independent justification. 
This means, that an essential motivation for split supersymmetry, gauge unification, is only 
accidentally realized. 


\section{Hierarchical D-term inflation}

In hierarchical supersymmetry breaking of the type discussed here 
it seems an intriguing suggestion to relate the 
mass scale of the heavy scalars to cosmology. In our model, the potential energy is generated at 
the conventional supersymmetry breaking scale $(m_{3/2} M_{\rm Pl})^2$, and with standard values 
comes out about $(10^{10-13}\, {\rm GeV})^4$. This is the scale of the individual contributions 
of the F- and D-terms to the full potential, and only the fine tuning of the cosmological constant 
leads to a cancellation. It is now very tempting to identify the vacuum energy of these individual 
components with the vacuum energy that drives inflation, by undoing the fine  tuning. 
A possible scenario is very easily illustrated 
along the lines of D-term inflation \cite{Binetruy:1996xj}. 
Roughly speaking, the only required modification of the model
we used so far, with the MSSM extended by one or many extra $U(1)_a$ gauge factors, plus a pair of charged fields 
$\phi^\pm$, is to promote the mass term (\ref{supo}) in the superspotential to a dynamical 
field $\varphi$, neutral under $U(1)_a$, which plays the role of the inflaton. We now write 
\beqn 
W ~=~ \varphi \phi^+\phi^- + W_0 \ , 
\eeqn 
and use a canonically normalized K\"ahler potential, 
\beqn 
\ck ~=~ |\phi^+|^2 + |\phi^-|^2 + |\varphi|^2 + \ck_0\ . 
\eeqn 
With this one finds the scalar potential of the form 
\beqn 
\cv &=& \cv_{\rm hid} + e^{\kappa^2\ck} \Big[ |\varphi\phi^+|^2 + |\varphi\phi^-|^2 + |\phi^+\phi^-|^2 + 
 \kappa^4 \big( |\varphi|^2 + |\phi^+|^2 + |\phi^-|^2 \big) |W|^2 \non 
&& \hspace{1cm} + 3 \kappa^2 |\varphi\phi^+\phi^-|^2 
 - 3\kappa^2 |W_0|^2 \Big] + 
 \sum_a \frac{g_a^2}{2} \Big[ |\phi^+|^2 - |\phi^-|^2 + \xi_a \Big]^2 \ , 
\eeqn 
where the only negative contribution comes from $-3 \kappa^2 e^{\kappa^2\ck} |W_0|^2$. 
Above some threshold value for $\varphi$, the two charged fields 
$\phi^\pm$ are stabilized at the origin $\phi^\pm=0$. Their masses at zero are 
\beqn 
m^2_{\phi^\pm}(\phi^\pm=0) ~=~  \kappa^2 \cv_{\rm hid} +    e^{\kappa^2\ck_0}
[(1+\kappa^6|W_0|^2) |\varphi|^2  - 2\kappa^4 |W_0|^2] \pm \sum_a g_a^2 \xi_a \ , 
\eeqn 
which turns positive, when $\langle\varphi\rangle$ is large enough 
(but still well below the Planck scale). 
The potential simplifies to 
\beqn 
\cv (\phi^\pm=0) ~=~ \cv_{\rm hid} + \kappa^2 e^{\kappa^2\ck_0} |W_0|^2 [ \kappa^2 |\varphi|^2-3 ] 
 +  \sum_a \frac{g_a^2}{2} \xi_a^2 \ . 
\eeqn 
The inflationary slow-roll condition for the second derivative of the potential is \cite{book}
\beqn 
|\eta| ~=~ \left| \frac{\partial_\varphi^2 \cv }{\kappa^2\cv} \right| ~\ll~ 1\ , 
\eeqn 
where $|\eta|\ll 1$ numerically means $|\eta|\sim 0.01$ is acceptable. 
This implies that 
\beqn 
\frac{M_{\rm Pl}^2 \sum_a g_a^2 \xi_a^2}{e^{\kappa^2\ck} |W_0|^2} ~\gtrsim~ 100\ . 
\eeqn 
The first derivative is then automatically also small (with $\varphi \sim \sqrt{\xi_a}$), 
and inflation can be successful. 
This means that the very same D-term vacuum energy that is responsible for the large scalar masses 
can drive inflation, if either the D-terms are enhanced or the F-term vacuum energy is 
sufficiently suppressed during that period. 
Thus, during the de Sitter phase the relation between the Hubble parameter $H$ and
the energy density $\rho$, i.e. the relation $3M_{\rm Pl}^2 H^2 =\rho =\frac{1}{2} \dot\phi^2 +\cv$,
shows that the Hubble expansion in this phase is dominated by the FI-term 
\beqn 
3M_{\rm Pl}^2H^2 ~\sim~ \sum_a \frac{g_a^2}{2} \xi^2_a \ . 
\eeqn  
In summary, if in the phase of large $\varphi$ the vacuum energy of the D-terms is 
larger than that of the F-terms by a factor of about 100 or more, the D-term energy 
can drive inflation at a scale up to $10^{12-15}\, {\rm GeV}$, slightly above 
the mass scale of the heavy scalars. After the end of the slow roll period, $\varphi$ will 
eventually fall below the threshold value, and $\phi^\pm$ will form 
condensates themselves. This can then lead to a partial relaxation 
of the D-term vacuum energy, but a more elaborate model of the hidden sector would be needed to 
describe this phase transition properly. In the minimum, one may expect $\varphi$ to settle down to 
$\langle \varphi\rangle\sim \sqrt \xi\sim \sqrt{m_{3/2}M_{\rm Pl}}$ on dimensional grounds, which is compatible with 
small gaugino masses.\footnote{As 
mentioned earlier, we so far ignore here the problem of moduli stabilization, which is even more 
severe in the context of inflation with D-brane degrees of freedom, where already the correct choice 
of supersymmetric coordinates is a very subtle question \cite{Kachru:2003sx}. The FI-parameters 
depend on the moduli fields, and it will be necessary to stabilize the relevant fields 
(as for instance along the lines of \cite{Kachru:2003aw}) at a scale larger 
that the scale of inflation, or they would have to be considered as dynamical (see also 
the last reference of \cite{Binetruy:1996xj}).}


\section{Conclusion}

We have presented a model of supersymmetry breaking in the context of 
string and supergravity scenarios by inclusion of both F- and Fayet-Iliopoulos D-terms, arising from 
extra $U(1)$ factors in the gauge group. Such extra $U(1)$ 
gauge symmetries arise quite naturally in 
string based models. It was shown that scalars charged under the extra $U(1)$ gain 
large masses from the FI-terms, proportional to the charges of the respective scalar
fields under the extra $U(1)$. This leads generically to non-universal masses for
the heavy scalars. The cancellation of the vacuum energy puts an upper bound 
of $\sqrt{m_{3/2}M_{\rm Pl}}$ on the scalar masses, and thus also puts a bound on the 
FI-term $\xi_X$. The bound on $\xi_X$ can be met in  
heterotic string models only for $m_{3/2}$ close to $M_{\rm Pl}$, since there $\xi_X$
is scaled by $M_{\rm Pl}^2$. Thus, heterotic string
scenarios are not preferred from the vacuum energy constraint, when
$m_{3/2}=\co({\rm TeV})$. However, $m_{3/2}=\co({\rm TeV})$ and $\xi_X \sim m_{3/2}M_{\rm Pl}$,
i.e. of size $10^{10-13}\,{\rm GeV}$, could arise in orientifolds 
models which allow $\xi_X$ of a variable moduli-dependent size. 
The fact that the D-term contributions to the scalar masses depend on their $U(1)$ charges 
opens up the possibility of building a new class of models with some scalars (with vanishing 
$U(1)$ charges) light and others (with non-vanishing $U(1)$ charges) heavy,
while the gauginos and higgsinos gain masses only of electro-weak size. Further, 
the $\mu$ term is essentially unaffected by the FI contribution, and
can be of electro-weak size. We investigated two illustrative examples of models
with light and heavy scalars (Model I and Model II in Sec.~2) and showed that they lead to 
significantly different phenomenologies which could be tested at colliders and in
non-accelerator experiment. The class of models we have discussed here are 
different from the high scale supersymmetry models of Ref.~\cite{Arkani-Hamed:2004fb}, 
where all scalars and the gravitino are super heavy. However, which scalars 
are heavy and which are light is now a model-dependent question. A further interesting feature 
is the possibility that the vacuum energy responsible for generating heavy scalars  
may also drive inflation. An analysis of how this can come about was discussed 
in Sec.~4. It would be interesting
to investigate more explicit D-brane constructions to build models of the type advocated here.     


\section*{Acknowledgements}

B.~K.~would like to thank Angel Uranga for helpful advice. 
The work of B.~K.~was supported by the German Science Foundation (DFG) and in part by
funds provided by the U.S. Department of Energy (D.O.E.) under cooperative research agreement
$\#$DF-FC02-94ER40818. The work of P.~N.~was supported in part by
the U.S. National Science Foundation under the grant NSF-PHY-0139967. 


\end{document}